# BAK1 Gene Variation: the doubts remain

Michel E. Beleza Yamagishi[1]

**Laboratório de Bioinformática Aplicada**

**Embrapa Informática Agropecuária**

**Introduction**

Since Gottlieb et al. [2009] published some intriguing sequence differences between DNA obtained from abdominal aortic (AA) tissue and blood, the scientific community has been attempting to understand their findings. Dr. Gottlieb's group was primarily interested in performing case-control association analyses in abdominal aortic aneurysm (AAA) patients. However, instead of obtaining a significant association, they realized that both diseased and non-diseased AA tissues contained polymorphisms which were not observed in the matching blood samples. Dr. Hatchwell [2010] has proposed that the BAK1 gene variants were likely due to sequencing of a processed gene on chromosome 20. However, in response, Dr. Gottlieb and co-authors [2010] have argued that "some but not all of the sequence changes present in the BAK1 sequence of our abdominal aorta samples are also present in the chromosome 20 BAK1 sequence. However, all the AAA and AA cDNA samples are identical to each other and different from chromosome 20 BAK1 sequence at amino acids 2 and 145". I have been following this discussion because I have independently reached almost the same conclusion as Dr. Hatchwell did [Yamagishi, 2009], and, unfortunately, the response from Dr. Gottlieb and his co-authors seems to me to be unsatisfactory for the reasons listed below.

---

[1] E-mail: michel@cnptia.embrapa.br

## Sequence Inconsistencies

The core of their argument is that "some but not all of the sequence changes present in the BAK1 sequence of our abdominal aorta samples are also present in the chromosome 20 BAK1 sequence". Actually, there are 11 Single Nucleotide Polymorphisms (SNPs) as illustrated in Figure 1.

```
>ref|NT_011362.10|  Homo sapiens chromosome 20 genomic contig, GRCh37 reference primary assembly
Length=31409461

 Score = 1114 bits (603),  Expect = 0.0
 Identities = 625/636 (98%), Gaps = 0/636 (0%)
 Strand=Plus/Minus

Query  1        ATGGCTTCGGGGCAAGGCCCAGGTCCTCCCAGGCAGGAGTGCGGAGAGCCTGCCCTGCCC  60
Sbjct  1474709  .....C......................................................  1474650
Query  61       TCTGCTTCTGAGGAGCAGGTAGCCCAGGACACAGAGGAGGTTTTCCGCAGCTACGTTTTT  120
Sbjct  1474649  ............................................................  1474590
Query  121      TACCGCCATCAGCAGGAACAGGAGGCTGAAGGGGTGGCTGCCCCTGCCGACCCAGAGATG  180
Sbjct  1474589  ....A........................C..............................  1474530
Query  181      GTCACCTTACCTCTGCAACCTAGCAGCACCATGGGGCAGGTGGGACGGCAGCTCGCCATC  240
Sbjct  1474529  ............................................................  1474470
Query  241      ATCGGGGACGACATCAACCGACGCTATGACTCAGAGTTCCAGACCATGTTGCAGCACCTG  300
Sbjct  1474469  ..T.........................................................  1474410
Query  301      CAGCCCACGGCAGAGAATGCCTATGAGTACTTCACCAAGATTGCCACCAGCCTGTTTGAG  360
Sbjct  1474409  .............................................T..............  1474350
Query  361      AGTGGCATCAATTGGGGCCGTGTGGTGGCTCTTCTGGGCTTCGGCTACCGTCTGGCCCTA  420
Sbjct  1474349  .......................................A....................  1474290
Query  421      CACGTCTACCAGCATGGCCTGACTGGCTTCCTAGGCCAGGTGACCCGCTTCGTGGTCGAC  480
Sbjct  1474289  ...A.........G..................G.................T.....G...  1474230
Query  481      TTCATGCTGCATCACTGCATTGCCCGGTGGATTGCACAGAGGGGTGGCTGGGTGGCAGCC  540
Sbjct  1474229  ............................................................  1474170
Query  541      CTGAACTTGGGCAATGGTCCCATCCTGAACGTGCTGGTGGTTCTGGGTGTGGTTCTGTTG  600
Sbjct  1474169  ............................................................  1474110
Query  601      GGCCAGTTTGTGGTACGAAGATTCTTCAAATCATGA       636
Sbjct  1474109  ....................................      1474074
```

**Figure 1. Blast alignment of BAK1 gene on chromosome 6 and on chromosome 20.** The query sequence is the BAK1 gene on chromosome 6 and the Sbjct sequence is the BAK1 processed gene on chromosome 20. The eleven single nucleotide polymorphisms (SNPs) found in the BAK1 gene on chromosome 20 are clearly indicated. Observe that only the coding sequences are aligned.

However, they have picked only two previously not reported SNPs to support their argument. The problem is that the BAK1 sequence inferred from their Table 2 (third row) of their original paper does not seem to be the BAK1 refseq NM_001188.3 either. If they did not make multiple independent typographical errors, their BAK1 sequence differs from refseq NM_001188.3 in at least four codons as shown in Table 1. For instance, the 28$^{th}$ amino acid codon in their sequence is GTC which is mis-translated as ALA instead of VAL, while in sequence NM_001188.3 the codon is GCC (ALA) as is shown in Figure 1. The 42$^{nd}$ amino acid codon, CGC, is really problematic, for in refseq NM_001188.3 it is supposed to code for ARG, while in BAK1 on chromosome 20 the codon is CAC which codes for HIS. It occurs that in their Table 2 (rows 4 and 5, non-diseased AA and AAA) the codon CAA appears, which in fact codes for GLN. The 52$^{nd}$ amino acid codon is GTG (VAL), while in their Table 2 it is GCT (ALA), and, furthermore, codon GCC (ALA) is found both in non-diseased and AAA samples. Thus, some of their codons do not belong either to refseq NM_001188.3 or to the BAK1 on chromosome 20. The 103$^{rd}$ amino acid in their sequence is coded for by GCC, while it is coded for by ACG in refseq. Furthermore, there are also some inconsistencies, which are probably copy editing errors (Forsdyke - personal communication). For example, GCT codes for ALA and not VAL as appears in their Table 2.

The doubts remain. Their AAA and AA BAK1 sequence seems to represent a hybrid cDNA. Part of it belongs to the BAK1 processed gene on chromosome 20, another part belongs to the BAK1 gene on chromosome 6 and, what is more intriguing, there are some parts that do not belong to either, such as codons 28, 42, 52 and 103. It seems that in order to definitely clarify this issue, Gottlieb et al. should publish their full BAK1 sequence.

Table 1: Gottlieb's seq is the BAK1 sequence as inferred from Gottlieb et al. Table 2. Codon differences are in bold letters. The reported polymorphisms between the BAK1 gene on chromosome 6 and the BAK1 processed gene on chromosome 20 are shown as well. The main point is to demonstrate that Gottlieb's seq seems differ from refseq NM_001188 and BAK1 on chromosome 20 in at least four codons, namely codons 28, 42, 52 and 103. The amino acid inconsistencies are underlined.

| Amino acid # | 28 | 42 | 52 | 81 | 103 |
| --- | --- | --- | --- | --- | --- |
| Gottlieb's seq | G**T**C (ALA) | **CAG** (ARG) | G**CT** (**ALA**) | ATC (ILE) | **GCC** (**ALA**) |
| BAK1 gene (NM_001188) | GCC (ALA) | CGC (ARG) | GTG (VAL) | ATC (ILE) | ACG (THR) |
| BAK1 gene on chromosome 20 | GCC (ALA) | CAC (**HIS**) | GCG (ALA) | ATT (ILE) | ACG (THR) |
| Non-diseased AA | GTC (VAL) | CAA (GLN) | GCC (ALA) | ATT (ILE) | GCC (ALA) |
| AAA | GTC (VAL) | CAA (GLN) | GCC (ALA) | ATT (ILE) | GCC (ALA) |

## Primer Design

Gottlieb et al. used "BAK1-specific primers" to amplify AAA and AA samples; however, as correctly pointed out by Dr. Hatchwell, the Forward (CAGGCTGATCCCGTCCTCCACTGAG) and Reverse (GGGCACCCTTGGGAGTCATGATTTG) primers are just as likely to amplify the BAK1 gene on chromosome 20. There is no reason for the chromosome 6 region to amplify better than the chromosome 20 region, as the primers match perfectly to both, again demonstrating that their argument is not satisfactory.

It is true that Gottlieb et al., in their response, explicitly affirm that they have used intron-specific primers to amplify BAK1 exon 3 on AAA and AA tissue samples; however, from the text of the original article one was led to believe that they were using the non-specific primers. Thus, the primer design must be completely described, otherwise it is not possible to understand the results, much less the possible explanations.

Concerning the amplification of the matching blood samples, Gottlieb et al. indeed used BAK1 on chromosome 6 (exons) intron-specific primers; however, these samples are not in agreement with BAK1 refseq NM_001188.3 at least in amino acid codons 28, 42, 52 and 103 as shown in Table 1.

**Other Issues**

In their response, Gottlieb et al. [2010] said "we agree with Dr. Hatchell that BAK2 gene in chromosome 20 is a pseudogene"; however , Dr. Hatchwell did not state that the chromosome 20 homologue is a pseudogene, and therefore Gottlieb et al. misquoted him. In a personal communication, Dr. Hatchwell said that he deliberately avoided use of the term pseudogene, which implies an inactive gene. He rather described it as a processed (intronless) gene and, in fact, there is evidence that it is transcribed.

Finally, an experiment to determine the truth might consist of the use of intron-specific primers to the chromosome 6 BAK1 whole set of exons on AAA and AA samples. I believe that this is the definitive experiment.

**Acknowledgments**

We thank Roberto Herai for the preparation of the high resolution figure, José Andres Yunes who initially drew the matter to our attention and Donald Forsdyke for noting the errors in Table 2 of Gottlieb et al. We thank Dr. Hatchwell for in depth discussions.